\documentclass{llncs}

\renewcommand{\tt}{\usefont{OT1}{cmtt}{m}{n}\selectfont}

\usepackage{latexsym}
\usepackage{amssymb}
\usepackage{xspace}
\usepackage{epsfig}

\makeatletter
\newenvironment{prog}{\ifhmode\par\vspace{0.5ex}\fi
\setlength{\parindent}{3ex}
\setlength{\parskip}{0.0ex}
\obeylines\@vobeyspaces\tt}{\vspace{0.5ex}\noindent
}
\makeatother
\newcommand{\startprog}{\begin{prog}}
\newcommand{\stopprog}{\end{prog}\noindent}
\newcommand{\code}[1]{\mbox{\tt #1}}   
\newcommand{\ccode}[1]{``\mbox{\tt #1}''}  
\newcommand{\bs}{\char92} 
\newcommand{\us}{\char95} 

\newcommand{\startdef}{\begin{definition} \rm}
\newcommand{\startnameddef}[1]{\begin{definition}[#1] \rm}
\newcommand{\stopdef}{\leavevmode\hbox{ }\hfill$\Box$\smallskip \end{definition}}
\newcommand{\startlemma}{\begin{lemma} \sl}
\newcommand{\startnamedlemma}[1]{\begin{lemma}[#1] \sl}
\newcommand{\stoplemma}{\end{lemma}}
\newcommand{\starttheo}{\begin{theorem} \sl}
\newcommand{\startnamedtheo}[1]{\begin{theorem}[#1] \sl}
\newcommand{\stoptheo}{\end{theorem}}
\newcommand{\startcor}{\begin{cor} \sl}
\newcommand{\startnamedcor}[1]{\begin{cor}[#1] \sl}
\newcommand{\stopcor}{\end{cor}}
\newcommand{\startprop}{\begin{prop} \sl}
\newcommand{\startnamedprop}[1]{\begin{prop}[#1] \sl}
\newcommand{\stopprop}{\end{prop}}
\newcommand{\startex}{\begin{example}}
\newcommand{\startnamedex}[1]{\begin{example}[#1]}
\newcommand{\stopex}{\leavevmode\hbox{ }\hfill$\Box$\end{example}}
\newcommand{\startproof}{\begin{proof}}
\newcommand{\stopproof}{\end{proof}}

\newcommand{\EN}{\mbox{\it En}\xspace} 
\newcommand{\en}{\mbox{\it en}\xspace} 

\begin{document}
\sloppy

\title{Compiling ER Specifications into Declarative Programs%
  \thanks{This work was partially supported by the
   German Research Council (DFG) under grant Ha 2457/5-2.}}

\author{
Bernd Bra\ss{}el
\kern1em
Michael Hanus
\kern1em
Marion M\"uller
}
\institute{
Institut f\"ur Informatik, CAU Kiel, D-24098 Kiel, Germany. \\
\email{\{bbr|mh|mam\}@informatik.uni-kiel.de}
}

\maketitle

\begin{abstract}
This paper proposes an environment to support high-level database programming
in a declarative programming language.
In order to ensure safe database updates,
all access and update operations related to the database
are generated from high-level descriptions in the
entity-relationship (ER) model.
We propose a representation of ER diagrams
in the declarative language Curry so that
they can be constructed by various tools and
then translated into this representation.
Furthermore, we have implemented a compiler from
this representation into a Curry program that provides
access and update operations based on a high-level API
for database programming.
\end{abstract}

\section{Motivation}
\label{sec:motivation}

Many applications in the real world need databases to store the data
they process. Thus, programming languages for such applications
must also support some mechanism to organize the access to databases.
This can be done in a way that is largely independent on the
underlying programming language, e.g., by passing SQL statements
as strings to some database connection.
However, it is well known that such a loose coupling is the
source of security leaks, in particular, in web applications \cite{Huseby03}.
Thus, a tight connection or amalgamation of the database access
into the programming language should be preferred.

In principle, logic programming provides a natural framework
for connecting databases (e.g., see \cite{Das92,GallaireMinker78})
since relations stored in a relational database
can be considered as facts defining a predicate of a logic program.
Unfortunately, the well-developed theory in this area
is not accompanied by practical implementations.
For instance, distributions of Prolog implementations rarely come
with a standard interface to relational databases.
An exception is Ciao Prolog that has a persistence module \cite{CorreasEtAl04}
that allows the declaration of predicates where the facts
are persistently stored, e.g., in a relational database.
This module supports a simple method to query the relational database,
but updates are handled by predicates with side effects
and transactions are not explicitly supported.
A similar concept but with a clear separation between
queries, updates, and transactions has been proposed in
\cite{Hanus04JFLP} for the multi-paradigm declarative language
Curry \cite{Hanus97POPL,Hanus06Curry}.
This will be the basis for the current framework that provides an environment
for high-level programming with databases.
The objectives of this work are:
\begin{itemize}
\item The methods to access and update the database should be
expressed by language features rather than passing SQL strings around.
\item Queries to the database should be clearly separated from updates
that might change the outcome of queries.
\item Safe transactions, i.e., sequence of updates that keep some
integrity constraints, should be supported.
\item The necessary code for these operations should be derived
from specifications whenever possible in order to obtain more reliable
applications.
\end{itemize}
For this purpose, we define an API for database programming
in Curry that abstracts from the concrete methods to access a given database
by providing abstract operations for this purpose.
In particular, this API exploits the type system of Curry
in order to ensure a strict separation between queries and updates.
This is described in detail in Section~\ref{sec:db-programming}.
To specify the logical structure of the data to be stored in a database,
we use the entity-relationship (ER) model \cite{Chen76},
which is well established for this purpose.
In order to be largely independent of concrete specification tools,
we define a representation of ER diagrams in Curry
so that concrete ER specification tools can be connected
by defining a translator from the format used in these tools
into this Curry representation. This representation is described
in Section~\ref{sec:erd}.
Finally, we develop a compiler that translates an ER specification
into a Curry module that contains access and update operations
and operations to check integrity
constraints according to the ER specification.
The generated code uses the database API described 
in Section~\ref{sec:db-programming}.
The compilation method is sketched in Section~\ref{sec:erd2curry}.
Finally, Section~\ref{sec:conclusions} contains our conclusions.

\section{Database Programming in Curry}
\label{sec:db-programming}

We assume basic familiarity with
functional logic programming (see \cite{Hanus07ICLP} for a recent survey)
and Curry \cite{Hanus97POPL,Hanus06Curry} so that
we give in the following only a short sketch of the basic
concepts relevant for this paper.

Functional logic languages integrate the most important features
of functional and logic languages to provide a variety
of programming concepts to the programmer.
For instance, the concepts of demand-driven evaluation,
higher-order functions, and polymorphic typing from functional programming
are combined with logic programming features like
computing with partial information (logic variables),
constraint solving, and non-deterministic search for solutions.
This combination, supported by optimal evaluation strategies
\cite{AntoyEchahedHanus00JACM} and new design patterns
\cite{AntoyHanus02FLOPS}, leads to
better abstractions in application programs
such as implementing
graphical user interfaces \cite{Hanus00PADL},
programming dynamic web pages \cite{Hanus01PADL,Hanus06PPDP},
or access and manipulation of persistent data possibly stored
in databases \cite{Fischer05,Hanus04JFLP}.

As a concrete functional logic language, we use Curry in our framework
but it should be possible to apply the same ideas also to other
functional logic languages, e.g.,  TOY \cite{Lopez-FraguasSanchez-Hernandez99}.
{}From a syntactic point of view, a Curry program is a functional
program extended by the possible inclusion of free (logic)
variables in conditions and right-hand sides of defining rules.
Curry has a Haskell-like syntax \cite{PeytonJones03Haskell},
i.e., (type) variables and function names usually
start with lowercase letters and the names of type and data constructors
start with an uppercase letter. The application of $f$
to $e$ is denoted by juxtaposition (``$f~e$'').
A Curry \emph{program} consists of the definition of functions
and data types on which the functions operate.
Functions are first-class citizens as in Haskell and are evaluated lazily.
To provide the full power of logic programming,
functions can be called with partially instantiated arguments
and defined by conditional equations
with constraints in the conditions.
Function calls with free variables are evaluated by a possibly
nondeterministic instantiation of demanded arguments
(i.e., arguments whose values are necessary to decide the applicability
of a rule) to the required values in order to apply a rule.

\begin{example}\rm
\label{ex-conc}
The following Curry program defines the data types of
Boolean values, ``possible'' (maybe) values,
union of two types,
and polymorphic lists (first four lines)
and functions for computing the concatenation of lists and the last
element of a list:
\startprog
data Bool       = True    | False
data Maybe a    = Nothing | Just a
data Either a b = Left a  | Right b
data List a     = []      | a : List a
\smallskip
conc :: [a] -> [a] -> [a]
conc []     ys = ys
conc (x:xs) ys = x : conc xs ys
\smallskip
last :: [a] -> a
last xs | conc\,\,ys\,\,[x] =:= xs   = x   where x,ys free
\stopprog
The data type declarations define
\code{True} and \code{False} as the Boolean constants,
\code{Nothing} and \code{Just} as the constructors for possible values
(where \code{Nothing} is considered as no value),
\code{Left} and \code{Right} to inject values into
a union (\code{Either}) type,
and \code{[]} (empty list) and \code{:} (non-empty list) as the constructors
for polymorphic lists (\code{a} and \code{b} are type variables ranging over
all types and the type \ccode{List\,\,a} is usually written as \code{[a]}
for conformity with Haskell).
\end{example}
Curry also offers other standard features of
functional languages, like higher-order functions
(e.g., \ccode{\bs{}$x\,$->$\,e$} denotes an anonymous function
that assigns to each $x$ the value of $e$), modules,
or monadic I/O \cite{Wadler97}.
For instance, an operation of type \ccode{IO\,\,t} is an I/O action,
i.e., a computation that interacts with the ``external world''
and returns a value of type \code{t}.
Thus, purely declarative computations are distinguished from
I/O actions by their types so that they cannot be freely mixed.

Logic programming is supported by admitting function calls with
free variables (see \ccode{conc\,\,ys\,\,[x]} above)
and constraints in the condition of a defining rule.
Conditional program rules have the form $l~|~c~=~r$
specifying that $l$ is reducible to $r$ if the condition $c$ is satisfied
(see the rule defining \code{last} above).
A \emph{constraint} is any expression of the
built-in type \code{Success}.
For instance, the trivial constraint \code{success} is
an expression of type \code{Success}
that denotes the always satisfiable constraint.
\ccode{$c_1$\,\&\,$c_2$} denotes the \emph{concurrent conjunction}
of the constraints $c_1$ and $c_2$, i.e., this expression
is evaluated by proving both argument constraints concurrently.
An \emph{equational constraint} \code{$e_1$\,=:=\,$e_2$} is satisfiable if
both sides $e_1$ and $e_2$ are reducible to unifiable constructor terms.
Specific Curry systems also support more powerful
constraint structures, like arithmetic constraints on real numbers
or finite domain constraints
(e.g., PAKCS \cite{Hanus06PAKCS}).

Using functions instead of predicates has the advantage
that the information provided by functional dependencies
can be used to reduce the search space and evaluate goals
in an optimal way (e.g., shortest derivation sequences,
minimal solution sets, see \cite{AntoyEchahedHanus00JACM} for details).
However, there are also situations where a relational style
is preferable, e.g., for database applications as considered in this paper.
This style is supported by considering predicates
as functions with result type \code{Success}.
For instance, a predicate \code{isPrime} that is satisfied if the
argument (an integer number) is a prime can be modeled as a function
with type
\startprog
isPrime :: Int -> Success
\stopprog
The following rules define a few facts for this predicate:
\label{isprime-pred}
\startprog
isPrime 2 = success
isPrime 3 = success
isPrime 5 = success
isPrime 7 = success
\stopprog
Apart from syntactic differences,
any pure logic program has a direct correspondence to a Curry program.
For instance, a predicate \code{isPrimePair} that is satisfied
if the arguments are primes that differ by 2 can be defined
as follows:
\startprog
isPrimePair :: Int -> Int -> Success
isPrimePair x y = isPrime x \& isPrime y \& x+2 =:= y
\stopprog
In order to deal with information that is persistently stored
outside the program (e.g., in databases), \cite{Hanus04JFLP}
proposed the concept of dynamic predicates.
A \emph{dynamic predicate} is a predicate where the defining facts
(see \code{isPrime}) are not part of the program but stored outside.
Moreover, the defining facts can be modified (similarly to dynamic
predicates in Prolog).
In order to distinguish between definitions in a program (that do not
change over time) and dynamic entities, there is a distinguished type
\code{Dynamic} for the latter.\footnote{In contrast to Prolog,
where dynamic declarations are often used for efficiency purposes,
this separation is also necessary here due to the lazy evaluation
strategy which makes it difficult to estimate \emph{when} a particular
evaluation is performed. Thus, performing updates by implicit
side effects is not a good choice.}
For instance, in order to define a dynamic predicate \code{prime}
to store prime numbers whenever we compute them,
we provide the following definition in our program:
\startprog
prime :: Int -> Dynamic
prime dynamic
\stopprog
If the prime numbers should be persistently stored, one has to replace
the second line by
\startprog
prime persistent "$store$"
\stopprog
where $store$ specifies the storage mechanism, e.g., a directory
for a lightweight file-based implementation or a database specification
\cite{Fischer05}.

There are various primitives that deal with dynamic predicates.
First, there are combinators to construct complex queries from
basic dynamic predicates. For instance, the combinator
\startprog
(<>) :: Dynamic -> Dynamic -> Dynamic
\stopprog
joins two dynamic predicates, and the combinators
\startprog
(|>)  :: Dynamic -> Bool    -> Dynamic
(|\&>) :: Dynamic -> Success -> Dynamic
\stopprog
restrict a dynamic predicate with a Boolean condition or constraint,
respectively.
Since the operator \ccode{<>} binds stronger then \ccode{|>},
the expression
\startprog
prime x <> prime y |> x+2\,==\,y
\stopprog
specifies numbers \code{x} and \code{y} that are prime pairs.\footnote{%
Since the right argument of \ccode{|>} demands a Boolean value
rather than a constraint, we use the Boolean equality operator \ccode{==}
rather than the equational constraint \ccode{=:=} to compare the primes
\code{x} and \code{y}.}
On the one hand, such expressions can be translated into corresponding
SQL statements \cite{Fischer05} so that the programmer is freed
of dealing with details of SQL.
On the other hand, one can use all elements and libraries
of a universal programming language for database programming
due to its conceptual embedding in the programming language.

Since the contents of dynamic predicates can change over time,
one needs a careful concept of evaluating dynamic predicates
in order to keep the declarative style of programming.
For this purpose, we introduce a concept of queries
that are evaluated in the I/O monad, i.e., at particular points of time
in a computation.\footnote{Note that we only use the basic concept
of dynamic predicates from \cite{Hanus04JFLP}. The following
interface to deal with queries and transactions is new and more
abstract than the concepts described in \cite{Hanus04JFLP}.}
Conceptually, a \emph{query} is a method to compute solutions
w.r.t.\ dynamic predicates.
Depending on the number of requested solutions, there
are different operations to construct queries, e.g.,
\startprog
queryAll :: (a -> Dynamic) -> Query [a]
queryOne :: (a -> Dynamic) -> Query (Maybe a)
\stopprog
\code{queryAll} and \code{queryOne} construct queries
to compute all and one (if possible) solution to an abstraction
over dynamic predicates, respectively.
For instance,
\startprog
qPrimePairs :: Query [(Int,Int)]
qPrimePairs = queryAll (\bs{}(x,y) -> prime x <> prime y |> x+2\,==\,y)
\stopprog
is a query to compute all prime pairs.
In order to access the currently stored data, there is
an operation \code{runQ} to execute a query as an I/O action:
\startprog
runQ :: Query a -> IO a
\stopprog
For instance, executing the main expression \ccode{runQ qPrimePairs}
returns prime pairs that can be derived from the prime numbers
currently stored in the dynamic predicate \code{prime}.

In order to change the data stored in dynamic predicates,
there are operations to add and delete knowledge about dynamic predicates:
\startprog
addDB    :: Dynamic -> Transaction ()
deleteDB :: Dynamic -> Transaction ()
\stopprog
Typically, these operations are applied to single ground facts (since facts
with free variables cannot be persistently stored),
like \ccode{addDB (prime 13)} or \ccode{deleteDB (prime 4)}.
In order to embed these update operations into safe transactions,
the result type is \ccode{Transaction ()} (in contrast to the proposal
in \cite{Hanus04JFLP} where these updates are I/O actions).
A \emph{transaction} is basically a sequence of updates
that is completely executed or ignored (following the ACID principle
in databases).
Similarly to the monadic approach to I/O,
transactions also have a monadic structure so that
transactions can be sequentially composed
by a monadic bind operator:
\startprog
(|>>=) :: Transaction a -> (a\,\,->\,\,Transaction b) -> Transaction b
\stopprog
Thus, \ccode{t1 |>>= \bs{}x -> t2} is a transaction that
first executes transaction \code{t1}, which returns some result value
that is bound to the parameter \code{x} before executing transaction \code{t2}.
If the result of the first transaction is not relevant,
one can also use the specialized sequential composition \ccode{|>>}:
\startprog
(|>>) :: Transaction a -> Transaction b -> Transaction b
t1 |>> t2 = t1 |>>= \bs{}\us{} -> t2
\stopprog
A value can be mapped into a trivial transaction returning this value
by the usual return operator:
\startprog
returnT :: a -> Transaction a
\stopprog
In order to define a transaction that depend on some data
stored in a database, one can also embed a query into a transaction:
\startprog
getDB :: Query a -> Transaction a
\stopprog
For instance, the following expression exploits the standard higher-order
functions \code{map}, \code{foldr}, and \ccode{.} (function composition)
to define a transaction that deletes all known primes that are smaller
than \code{100}:
\startprog
getDB (queryAll (\bs{}i -> prime i |> i<100)) |>>=
foldr (|>>) (returnT ()) . map (deleteDB . prime)
\stopprog
Since such a sequential combination of transactions that are the
result of mapping a list of values into a list of transactions
frequently occurs, there is also a single function for this combination:
\startprog
mapT\us{} :: (a -> Transaction \us{}) -> [a] -> Transaction ()
mapT\us{} f = foldr (|>>) (returnT ()) . map f
\stopprog
To apply a transaction to the current database, there
is an operation \code{runT} that executes a given transaction
as an I/O action:
\startprog
runT :: Transaction a -> IO (Either a TError)
\stopprog
\code{runT} returns either the value computed by the successful
execution of the transaction or an error in case of a transaction failure.
The type \code{TError} of possible transaction errors
contains constructors for various kinds of errors, i.e., it is currently
defined as
\startprog
data TError = TError TErrorKind String
\smallskip
data TErrorKind = KeyNotExistsError | DuplicateKeyError
  | UniqueError | MinError | MaxError | UserDefinedError
\stopprog
but this type might be extended according to future requirements
(the string argument is intended to provide some details about
the reason of the error).
\code{UserDefinedError} is a general error that could be raised
by the application program whereas the other alternatives
are typical errors due to unsatisfied integrity constraints.
An error is raised inside a transaction by the operation
\startprog
errorT :: TError -> Transaction a
\stopprog
where the specialization
\startprog
failT :: String -> Transaction a
failT s = errorT (TError UserDefinedError s)
\stopprog
is useful to raise user-defined transaction errors.
If an error is raised inside a transaction, the transaction
is aborted, i.e., the transaction monad
satisfies the laws
\begin{center}
\begin{tabular}{r@{~~$=$~~}l}
$\code{errorT}~e ~\code{|>>=}~ t$ & \code{errorT~$e$} \\
$t ~\code{|>>=}~ \code{\bs{}$x$~->~errorT}~e$ & \code{errorT~$e$} \\
$\code{runT}~\code{(errorT$~e$)}$ & \code{return~(Right$~e$)}
\end{tabular}
\end{center}
Thus, the changes to the database performed in a transaction
that raises an error are not visible.

There are a few further useful operations on transactions
that are not relevant for this paper so that we omit them.
We summarize the important features of this abstract programming model
for databases:
\begin{itemize}
\item
Persistent data is represented in the application program
as language entities (i.e., dynamic predicates) so that
one can use all features of the underlying programming language
(e.g., recursion, higher-order functions, deduction)
for programming with this data.
\item
There is a clear separation between the data access (i.e., queries)
and updates that can influence the results of accessing data.
Thus, queries are purely declarative and are applied to
the actual state of the database when their results are required.
\item
Transactions, i.e., database updates, can be constructed
from a few primitive elements by specific combinators.
Transactions are conceptually executed as an atomic action
on the database. Transactions can be sequentially composed
but nested transactions are excluded due to the type system
(this feature is intended since nested transactions are usually
not supported in databases).
\end{itemize}
All features for database programming are summarized in
a specific \code{Database} library\footnote{%
\tt\url{http://www.informatik.uni-kiel.de/\char126pakcs/lib/CDOC/Database.html}}
so that they can be simply used
in the application program by importing it.
This will be the basis to generate higher-level code from
entity-relationship diagrams that are described in the following.

\section{Entity-Relationship Diagrams}
\label{sec:erd}

The entity-relationship model is a framework to specify
the structure and specific constraints of data stored in a database.
It uses a graphical notation, called entity-relationship diagrams (ERDs)
to visualize the conceptual model.
In this framework, the part of the world that is interesting for
the application is modeled by entities that have attributes
and relationships between the entities.
The relationships have cardinality constraints
that must be satisfied in each valid state of the database, e.g.,
after each transaction.

\begin{figure*}[t]
\begin{center}
  \epsfig{file=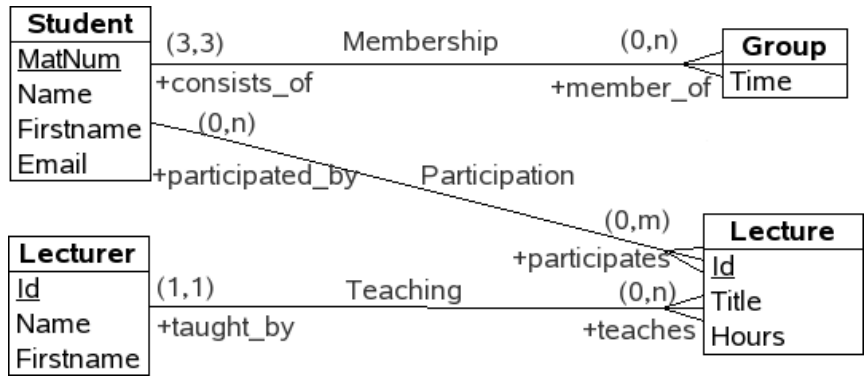,scale=0.6}
\end{center}
\caption{A simple entity-relationship diagram for university lectures}
\label{fig:erd}
\end{figure*}

There are various tools to support the data modeling process
with ERDs. In our framework we want to use some tool
to develop specific ERDs from which the necessary program code
based on the \code{Database} library described in the previous section
can be automatically generated.
In order to become largely independent of a concrete tool,
we define a representation of ERDs in Curry
so that a concrete ERD tool can be applied in this framework
by implementing a translator from the tool format into our representation.
In our concrete implementation, we have used the free software tool
Umbrello UML Modeller\footnote{\tt\url{http://uml.sourceforge.net}},
a UML tool part of KDE that also supports ERDs.
Figure~\ref{fig:erd} shows an example ERD constructed with this tool.
The developed ERDs are stored in XML files
in XMI (XML Metadata Interchange) format, a format for
the exchange of UML models.
Thus, it is a standard XML transformation task
to translate the Umbrello format into our ERD format.

The representation of ERDs as data types in Curry is straightforward.
A complete ERD consists of a name (that is later used as the module name
for the generated code) and lists of entities and relationships:
\startprog
data ERD = ERD String [Entity] [Relationship]
\stopprog
An entity has a name and a list of attributes, where
each attribute has a name, a domain, and specifications
about its key and null value property:
\startprog
data Entity = Entity String [Attribute]
\smallskip
data Attribute = Attribute String Domain Key Null
\smallskip
data Key = NoKey | PKey | Unique
\smallskip
type Null = Bool
\smallskip
data Domain = IntDom             (Maybe Int)
            | FloatDom           (Maybe Float)
            | CharDom            (Maybe Char)
            | StringDom          (Maybe String)
            | BoolDom            (Maybe Bool)
            | DateDom            (Maybe ClockTime)
            | UserDefined String (Maybe String)
            | KeyDom String   -- later used for foreign keys
\stopprog
Thus, each attribute is part of a primary key (\code{PKey}),
unique (\code{Unique}), or not a key (\code{NoKey}).
Furthermore, it is allowed that specific attributes can have null values, i.e.,
can be undefined. The domain of each attribute is one of the standard
domains or some user-defined type. In the latter case, the
first argument of the constructor \code{UserDefined}
is the qualified type name used in the Curry application program
(note that the \code{Database} library is able to handle
complex types by mapping them into standard SQL types \cite{Fischer05}).
For each kind of domain, one can also have a default value
(modeled by the \code{Maybe} type in Curry).
The constructor \code{KeyDom} is not necessary to represent ERDs
but will be later used to transform ERDs into relational database schema.

Finally, each relationship has a name and a list of connections
to entities (\code{REnd}), where each connection has the name
of the connected entity, the role name of this connection,
and its cardinality as arguments:
\startprog
data Relationship = Relationship String [REnd]
\smallskip
data REnd = REnd String String Cardinality
\smallskip
data Cardinality = Exactly Int | Range Int (Maybe Int)
\stopprog
The cardinality is either a fixed integer or a range between two
integers (where \code{Nothing} as the upper bound represents an
arbitrary cardinality).
For instance, the simple-complex (1:n) relationship \code{Teaching}
in Figure~\ref{fig:erd} can be represented by
the term
\startprog
Relationship "Teaching"
             [REnd "Lecturer" "taught\us{}by" (Exactly 1),
              REnd "Lecture" "teaches" (Range 0 Nothing)]
\stopprog

\section{Compiling ER Diagrams into Curry Programs}
\label{sec:erd2curry}

This section describes the transformation of ERDs into executable
Curry code. This transformation is done in the following order:
\begin{enumerate}
\item Translate an ERD into an \code{ERD} term.
\item Represent the relationships occurring in an \code{ERD} term
as entities.
\item Map all entities into corresponding Curry code
based on the \code{Database} library.
\end{enumerate}
The first step depends on the format used in the ERD tool.
As mentioned above, we have implemented a translation from
the XMI format used by the Umbrello UML Modeller into \code{ERD} terms.
This part is relatively easy thanks to the presence of XML processing tools.

\subsection{Transforming ERDs}

The second step is necessary since the relational model
supports only relations (i.e., database tables).
Thus, entities as well as relationships must be mapped into relations.
The mapping of entities into relations is straightforward
by using the entity name as the name of the relation
and the attribute names as column names.
The mapping of relationships is more subtle.
In principle, each relationship can be mapped into a corresponding relation.
However, this simple approach might cause the creation of many relations
or database tables. In order to reduce them, it is sometimes
better to represent specific relations as foreign keys,
i.e., to store the key of entity $e_1$ referred by a relationship
between $e_1$ and $e_2$ in entity $e_2$.
Whether or not this is possible depends on the kind of the relation.
The different cases will be discussed next.
Note that the representation of relationships as relations
causes also various integrity constraints to be satisfied.
For instance, if an entity has an attribute which contains a foreign key,
the value of this attribute must be either null or an existing key
in the corresponding relation. Furthermore, the various cardinalities
of each relationship must be satisfied.
Ideally, each transaction should ensure that the integrity constraints
are valid after finishing the transaction.

Now we discuss the representation of the various kinds of relationships
in the ER model.
For the sake of simplicity, we assume that each relationship
contains two ends, i.e., two roles with cardinality ranges $(min,max)$
so that we can characterize each relationship by their
related cardinalities $(min_A,max_A):(min_B,max_B)$ between
entities $A$ and $B$ (where $max_i$ is either a natural number
greater than $min_i$ or $\infty$, $i \in \{A,B\}$).
\label{sec:impl-relationships}
\begin{description}
\item[Simple-simple (1:1) relations:] 
This case covers all situations where each cardinality is at most one.
In the case $(0,1):(1,1)$, the key of entity $B$ is added
as an attribute to entity $A$ containing a foreign key
since there must be exactly one $B$ entity for each $A$ entity.
Furthermore, this attribute is \code{Unique} to ensure the uniqueness
of the inverse relation.
The case $(0,1):(0,1)$ can be similarly treated except that null values
are allowed for the foreign key.
\item[Simple-complex (1:n) relations:] 
In the case $(0,1):(min_B,max_B)$, the key of entity $A$ is added
as a foreign key (possibly null) to each $B$ entity.
If $min_B>0$ or $max_B \not= \infty$, the integrity constraints
for the right number of occurrences must be checked by each database update.
The case $(1,1):(0,max_B)$ is similarly implemented except
that null values for the foreign key are not allowed.
\item[Complex-complex (n:m) relations:] 
In this case a new relation representing this relationship is introduced.
The new relation is connected to entities $A$ and $B$ by two new
relationships of the previous kinds.
\end{description}
Note that we have not considered relationships where
both minimal cardinalities are greater than zero.
This case is excluded by our framework
(and rarely occurs in practical data models)
since it causes difficulties when creating new entities of type $A$ or $B$.
Since each entity requires a relation to an existing entity of the other type
and vice versa, it is not possible to create the new entities
independently. Thus, both entities must be created and connected
in one transaction which requires specific complex transactions.
Therefore, we do not support this in our code generation.
If such relations are required in an application (e.g.,
cyclic relationships), then the necessary code must be directly
written with the primitives of the \code{Database} library.

Based on this case distinction, the second step of our compiler
maps an \code{ERD} term into a new \code{ERD} term
where foreign keys are added to entities and new
entities are introduced to represent complex-complex relations.
Furthermore, each entity is extended with an internal primary key
to simplify the access to each entity by a unique scheme.

\subsection{Code Generation for ERDs}

After the mapping of entities and relationships into relations
as described above, we can generate the concrete program code
to organize the database access and update.
As already mentioned, we base the generated code on the
functionality provided by the library \code{Database}
described in Section~\ref{sec:db-programming}.
The schemas for the generated code are sketched in this section.
We use the notation \EN for the name of an entity (which starts
by convention with an uppercase letter) and $en$ for the same
name where the first letter is lowercase (this is necessary
due to the convention in Curry that data constructors and functions
start with uppercase and lowercase letters, respectively).

The first elements of the generated code are data types to
represent relations. For each entity \EN with attributes of types
$at_1,\ldots,at_n$, we generate the following two type definitions:
\startprog
data \EN = \EN Key $at_1$\ldots$at_n$
\smallskip
data \EN{}Key = \EN{}Key Key
\stopprog
\code{Key} is the type of all internal keys for entities.
Currently, it is identical to \code{Int}.
Thus, each entity structure contains an internal key for its
unique identification. The specific type \code{\EN{}Key}
is later used to distinguish the keys for different entities
by their types, i.e., to exploit the type system of Curry
to avoid confusion between the various keys.
For each relation that has been introduced for a
complex-complex relationship (see above),
a similar type definition is introduced except that it does not have
an internal key but only the keys of the connected entities as arguments.
Note that only the names of the types are exported but not their
internal structure (i.e., they are \emph{abstract} data types
for the application program).
This ensures that the application program cannot manipulate
the internal keys. The manipulation of attributes is possible
by explicit getter and setter functions that are described next.

In order to access or modify the attributes of an entity,
we generate corresponding functions where we use the attribute names
of the ERD for the names of the functions.
If entity \EN has an attribute $A_i$ of type $at_i$ ($i=1,\ldots,n$),
we generate the following getter and setter functions and a function
to access the key of the entity:
\startprog
\en{}$A_i$ :: \EN -> $at_i$
\en{}$A_i$ (\EN \us{} \ldots $x_i$ \ldots \us) = $x_i$
\smallskip
set\EN{}$A_i$ :: \EN -> $at_i$ -> \EN
set\EN{}$A_i$ (\EN $x_1$ \ldots \us{} \ldots $x_n$) $x_i$ = \EN $x_1$ \ldots $x_i$ \ldots $x_n$
\smallskip
\en{}Key :: \EN -> \EN{}Key
\en{}Key (\EN k \us{} \ldots \us) = \EN{}Key k
\stopprog
As described in Section~\ref{sec:db-programming},
data can be persistently stored by putting them into a dynamic predicate.
Thus, we define for each entity \EN a dynamic predicate
\startprog
\en{}Entry :: \EN -> Dynamic
\en{}Entry persistent "\ldots"
\stopprog
Since the manipulation of all persistent data should be done
by safe operations, this dynamic predicate is not exported.
Instead, a dynamic predicate \en is exported that associates
a key with the data so that an access is only possible to data
with an existing key:
\startprog
\en :: \EN{}Key -> \EN -> Dynamic
\en key obj | key\,=:=\,\en{}Key obj = \en{}Entry obj
\stopprog
Although these operations seem to be standard functions,
the use of a functional logic language is important here.
For instance, the access to an entity with a given key \code{k}
can be done by solving the goal \ccode{\en k o} where \code{o}
is a free variable that will be bound to the concrete instance of the entity.

For each role with name $rn$ specified in an ERD, we also generate
a dynamic predicate of type
\startprog
$rn$ :: $\EN_1$Key -> $\EN_2$Key -> Dynamic
\stopprog
where $\EN_1$ and $\EN_2$ are the entities related by this role.
The implementation of these predicates depend on the kind of
relationship according to their implementation as discussed
in Section~\ref{sec:impl-relationships}.
Since complex-complex relationships are implemented as relations,
i.e., persistent predicates (that are only internal and not exported),
the corresponding roles can be directly mapped to these.
Simple-simple and simple-complex relationships are implemented
by foreign keys in the corresponding entities.
Thus, their roles are implemented by accessing these keys.
We omit the code details that depend on the different
cases already discussed in Section~\ref{sec:impl-relationships}.

Based on these basic implementations of entities and relationships,
we also generate code for transactions to manipulate the data
and check the integrity constraints specified by the relationships
of an ERD.
In order to access an entity with a specific key,
there is a generic function that delivers this entity
in a transaction or raises a transaction error if there is
no entry with this key:
\startprog
getEntry :: k -> (k -> a -> Dynamic) -> Transaction a
getEntry key pred =
  getDB (queryOne (\bs{}info -> pred key info)) |>>=
  maybe (errorT (KeyNotExistsError "no entry for\ldots"))
        returnT
\stopprog
This internal function is specialized to an exported function for
each entity:
\startprog
get\EN :: \EN{}Key -> Transaction \EN
get\EN key = getEntry key \en
\stopprog
In order to insert new entities, there is a ``new'' transaction
for each entity. If the ERD specifies no relationship for this
entity with a minimum greater than zero, there is no need
to provide related entities so that the transaction has the
following structure (if \EN has attributes of types $at_1,\ldots,at_n$):
\startprog
new\EN :: $at_1$ -> $\cdots$ -> $at_n$ -> Transaction \EN
new\EN $a_1$ \ldots $a_n$ = $check_1$ |>> \ldots |>> $check_n$ |>> newEntry \ldots
\stopprog
Here, $check_i$ are the various integrity checks (e.g., uniqueness checks
for attributes specified as \code{Unique}) and \code{newEntry} is
a generic operation (similarly to \code{getEntry}) to insert a new entity.
If attribute $A_i$ has a default value or null values are allowed
for it, the type $at_i$ is replaced by \code{Maybe\,\,$at_i$} in \code{new\EN}.
If there are relationships for this entity with a minimum greater than zero,
than the keys (in general, a list of keys) must be also provided
as parameters to \code{new\EN}. The same holds for the ``new'' operations
generated for each complex-complex relationship.
For instance, the new operation for lectures according to
the ERD in Figure~\ref{fig:erd} has the following type:
\startprog
newLecture :: LecturerKey -> Int -> String -> Maybe Int
              -> Transaction Lecture
\stopprog
The first argument is the key of the lecturer required by the
relationship \code{Teaching}, and the further arguments
are the values of the \code{Id}, \code{Title} and \code{Hours} attributes
(where the attribute \code{Hours} has a default value
so that the argument is optional).

Similarly to \code{new\EN}, we provide also operations to update
existing entities. These operations have the following structure:
\startprog
update\EN :: \EN -> Transaction ()
update\EN e = $check_1$ |>> \ldots |>> $check_n$ |>> updateEntry \ldots
\stopprog
Again, the various integrity constraints must be checked
before an update is finally performed.
In order to get an impression of the kind of integrity constraints,
we discuss a few checks in the following.

For instance, if an attribute of an entity is \code{Unique},
this property must be checked before a new instance of the entity
is inserted. For this purpose, there is a generic transaction
\startprog
unique :: a -> (b -> a) -> (b -> Dynamic) -> Transaction ()
\stopprog
where the first argument is the attribute value, the second argument
is a getter function for this attribute, and the third argument
is the dynamic predicate representing this entity, i.e., a typical
call to check the uniqueness of the new value $a_i$
for attribute $A_i$ of entity \EN is \code{(unique $a_i$ \en{}$A_i$ \EN)}.
This transaction raises a \code{UniqueError} if an instance with
this attribute value already exists.

If an entity contains a foreign key, each update must check
the existence of this foreign key.
This is the purpose of the generic transaction
\startprog
existsDBKey :: k -> (a -> k) -> (a -> Dynamic) -> Transaction ()
\stopprog
where the arguments are the foreign key, a getter function (\code{\en{}Key})
for the key in the foreign entity and the dynamic predicate of
the foreign entity. If the key does not exist,
a \code{KeyNotExistsError} is raised.
Furthermore, there are generic transactions to check minimum
and maximum cardinalities for relationships and lists of foreign keys
that can raise the transaction errors \code{MinError}, \code{MaxError},
or \code{DuplicateKeyError}.
For each \code{new} and \code{update} operation generated
by our compiler, the necessary integrity checks are inserted
based on the ER specification.

Our framework does not provide delete operations.
The motivation for this is that safe delete operations
require the update of all other entities where this entity could
occur as a key. Thus, a simple delete could cause many implicit changes
that are difficult to overlook.
It might be better to provide only the deletion of single entities
followed by a global consistency check (discussed below).
A solution to this problem is left as future work.

Even if our generated transactions ensure the integrity of the
affected relations, it is sometimes useful to provide
a global consistency check that is regularly applied to all data.
This could be necessary if unsafe delete operations are performed,
or the database is modified by programs that do not use the safe
interface but directly accesses the data.
For this purpose, we also generate a global consistency test
that checks all persistent data w.r.t.\ the ER model.
If $E_1,\ldots,E_n$ are all entities (including the implicit
entities for complex-complex relations) derived from the given ERD,
the global consistency test is defined by
\startprog
checkAllData :: Transaction ()
checkAllData = check$E_1$ |>> \ldots |>> check$E_n$
\stopprog
The consistency test for each entity \EN is defined by
\startprog
check\EN :: Transaction ()
check\EN = getDB (queryAll \en{}Entry) |>>= mapT\us{} check\EN{}Entry
\smallskip
check\EN{}Entry :: \EN -> Transaction ()
check\EN{}Entry e = $check_1$ |>> \ldots |>> $check_n$
\stopprog
where the tests $check_i$ are similar to the ones used
in new and update operations that raise transaction errors
in case of unsatisfied integrity constraints.

\section{Conclusions}
\label{sec:conclusions}

We have presented a framework to compile conceptual data models
specified as entity-relationship diagrams into executable
code for database programming in Curry.
This compilation is done in three phases: translate the specific
ERD format into a tool-independent representation,
transform the relationships into relations according to their complexity,
and generate code for the safe access and update of the data.

Due to the importance of ERDs to design conceptual data models,
there are also other tools with similar objectives.
Most existing tools support only the generation
of SQL code, like the free software tools
DB-Main\footnote{\tt\url{http://www.db-main.be}}
or DBDesigner4\footnote{\tt\url{http://www.fabforce.net/dbdesigner4}}.
The main motivation for our development was the seamless embedding
of database programming in a declarative programming language
and the use of existing specification methods like ERDs
as the basis to generate most of the necessary code
required by the application programs.
The advantages of our framework are:
\begin{itemize}
\item The application programmer must only specify the data model
in a high-level format (ERDs) and all necessary code for dealing
with data in this model is generated.
\item The interface used by the application programs is type safe,
i.e., the types specified in the ERD are mapped into types
of the programming language so that ill-typed data cannot be constructed.
\item Updates to the database are supported as transactions
that automatically checks all integrity constraints specified in the ERD.
\item Checks for all integrity constraints are derived from the ERD
for individual tables and the complete database so that
they can be periodically applied to verify the integrity
of the current state of the database.
\item The generated code is based on an abstract interface for
database programming so that it is readable and well structured.
Thus, it can be easily modified and adapted to new requirements.
For instance, integrity constraints not expressible in ERDs
can be easily added to individual update operations, or
specific deletion operations can be inserted in the generated module.
\end{itemize}
For future work we intend to increase the functionality
of our framework, e.g., to extend ERDs by allowing the specification of
more complex integrity constraints or attributes for relations,
which is supported by some ER tools,
or to provide also delete operations for particular entities.
Finally, it could be also interesting to generate
access and update operations for existing databases
by analyzing their data model.
Although this is an issue different from our framework,
one can reuse the API described in Section~\ref{sec:db-programming}
and some other techniques of this paper for such a purpose.


\begin{thebibliography}{10}

\bibitem{AntoyEchahedHanus00JACM}
S.~Antoy, R.~Echahed, and M.~Hanus.
\newblock A Needed Narrowing Strategy.
\newblock {\em Journal of the ACM}, Vol.~47, No.~4, pp. 776--822, 2000.

\bibitem{AntoyHanus02FLOPS}
S.~Antoy and M.~Hanus.
\newblock Functional Logic Design Patterns.
\newblock In {\em Proc.\ of the 6th International Symposium on Functional and
  Logic Programming (FLOPS 2002)}, pp. 67--87. Springer LNCS 2441, 2002.

\bibitem{Chen76}
P.~P.-S. Chen.
\newblock The Entity-Relationship Model---Toward a Unified View of Data.
\newblock {\em ACM Transactions on Database Systems}, Vol.~1, No.~1, pp. 9--36,
  1976.

\bibitem{CorreasEtAl04}
J.~Correas, J.M. G{\'o}mez, M.~Carro, D.~Cabeza, and M.~Hermenegildo.
\newblock A Generic Persistence Model for (C)LP Systems (and Two Useful
  Implementations).
\newblock In {\em Proc. of the Sixth International Symposium on Practical
  Aspects of Declarative Languages (PADL'04)}, pp. 104--119. Springer LNCS
  3057, 2004.

\bibitem{Das92}
S.K. Das.
\newblock {\em Deductive Databases and Logic Programming}.
\newblock Addison-Wesley, 1992.

\bibitem{Fischer05}
S.~Fischer.
\newblock A Functional Logic Database Library.
\newblock In {\em Proc. of the ACM SIGPLAN 2005 Workshop on Curry and
  Functional Logic Programming (WCFLP 2005)}, pp. 54--59. ACM Press, 2005.

\bibitem{GallaireMinker78}
H.~Gallaire and J.~Minker, editors.
\newblock {\em Logic and Databases}, New York, 1978. Plenum Press.

\bibitem{Hanus97POPL}
M.~Hanus.
\newblock A Unified Computation Model for Functional and Logic Programming.
\newblock In {\em Proc.\ of the 24th ACM Symposium on Principles of Programming
  Languages (Paris)}, pp. 80--93, 1997.

\bibitem{Hanus00PADL}
M.~Hanus.
\newblock A Functional Logic Programming Approach to Graphical User Interfaces.
\newblock In {\em International Workshop on Practical Aspects of Declarative
  Languages (PADL'00)}, pp. 47--62. Springer LNCS 1753, 2000.

\bibitem{Hanus01PADL}
M.~Hanus.
\newblock High-Level Server Side Web Scripting in Curry.
\newblock In {\em Proc.\ of the Third International Symposium on Practical
  Aspects of Declarative Languages (PADL'01)}, pp. 76--92. Springer LNCS 1990,
  2001.

\bibitem{Hanus04JFLP}
M.~Hanus.
\newblock Dynamic Predicates in Functional Logic Programs.
\newblock {\em Journal of Functional and Logic Programming}, Vol.~2004, No.~5,
  2004.

\bibitem{Hanus06PPDP}
M.~Hanus.
\newblock Type-Oriented Construction of Web User Interfaces.
\newblock In {\em Proceedings of the 8th ACM SIGPLAN International Conference
  on Principles and Practice of Declarative Programming (PPDP'06)}, pp. 27--38.
  ACM Press, 2006.

\bibitem{Hanus07ICLP}
M.~Hanus.
\newblock Multi-paradigm Declarative Languages.
\newblock In {\em Proceedings of the International Conference on Logic
  Programming (ICLP 2007)}, pp. 45--75. Springer LNCS 4670, 2007.

\bibitem{Hanus06PAKCS}
M.~Hanus, S.~Antoy, B.~Bra{\ss}el, M.~Engelke, K.~H{\"o}ppner, J.~Koj,
  P.~Niederau, R.~Sadre, and F.~Steiner.
\newblock {PAKCS}: The {P}ortland {A}achen {K}iel {C}urry {S}ystem.
\newblock Available at \verb+http://www.informatik.uni-kiel.de/~pakcs/+, 2006.

\bibitem{Hanus06Curry}
M.~Hanus~(ed.).
\newblock Curry: An Integrated Functional Logic Language (Vers.\ 0.8.2).
\newblock Available at \verb+http://www.informatik.uni-kiel.de/~curry+, 2006.

\bibitem{Huseby03}
S.H. Huseby.
\newblock {\em Innocent Code: A Security Wake-Up Call for Web Programmers}.
\newblock Wiley, 2003.

\bibitem{Lopez-FraguasSanchez-Hernandez99}
F.~L\'opez-Fraguas and J.~S\'anchez-Hern\'andez.
\newblock {TOY}: A Multiparadigm Declarative System.
\newblock In {\em Proc. of RTA'99}, pp. 244--247. Springer LNCS 1631, 1999.

\bibitem{PeytonJones03Haskell}
S.~Peyton~Jones, editor.
\newblock {\em Haskell 98 Language and Libraries---The Revised Report}.
\newblock Cambridge University Press, 2003.

\bibitem{Wadler97}
P.~Wadler.
\newblock How to Declare an Imperative.
\newblock {\em ACM Computing Surveys}, Vol.~29, No.~3, pp. 240--263, 1997.

\end{thebibliography}
\end{document}